\documentclass{article}

\usepackage{arxiv}

\usepackage[utf8]{inputenc} 
\usepackage[T1]{fontenc}    
\usepackage{hyperref}       
\usepackage{url}            
\usepackage{booktabs}       
\usepackage{amsfonts}       
\usepackage{nicefrac}       
\usepackage{microtype}      
\usepackage{lipsum}		
\usepackage{graphicx}
\usepackage{doi}

\usepackage{amsmath} 
\usepackage{siunitx} 

\title{Multiscale analysis of crystal defect formation in rapid solidification of pure aluminium and aluminium-copper alloys}

\author{
Tatu Pinomaa$^{1}$, Matti Lindroos$^{1}$, Paul Jreidini$^2$,  Matias Haapalehto$^{1}$,  \\
\textbf{Kais Ammar$^{3}$, Lei Wang$^{4}$, Samuel Forest$^3$, Nikolas Provatas$^{2}$, and Anssi Laukkanen$^{1}$}
\\
\\
$^{1}$ICME group$,$ VTT Technical Research Centre of Finland Ltd$,$ Espoo, Finland,\\
$^{2}$Department of Physics and Centre for the Physics of Materials, McGill University, Montreal, Canada.\\
$^{3}$MINES ParisTech, PSL University, MAT - Centre des matériaux, Evry, France.\\
$^4$Federal Institute for Materials Research and Testing (BAM), Berlin, Germany.
}

\renewcommand{\shorttitle}{}

\hypersetup{
pdftitle={A template for the arxiv style},
pdfsubject={q-bio.NC, q-bio.QM},
pdfauthor={David S.~Hippocampus, Elias D.~Striatum},
pdfkeywords={First keyword, Second keyword, More},
}

\begin{document}
\maketitle

\begin{abstract}
Rapid solidification leads to unique microstructural features, where a less studied topic is the formation of various crystalline defects, including high dislocation densities, as well as gradients and splitting of the crystalline orientation. As these defects critically affect the material's mechanical properties and performance features, it is important to understand the defect formation mechanisms, and how they depend on the solidification conditions and alloying. 
To illuminate the formation mechanisms of the rapid solidification induced crystalline defects, we conduct a multiscale modeling analysis consisting of bond-order potential based molecular dynamics (MD), phase field crystal based amplitude expansion (PFC-AE) simulations, and sequentially coupled phase field -- crystal plasticity (PF--CP) simulations. The resulting dislocation densities are quantified and compared to past experiments.
The atomistic approaches (MD, PFC) can be used to calibrate continuum level crystal plasticity models, and the framework adds mechanistic insights arising from the multiscale analysis.
%
%

\end{abstract}


\section{Introduction}



Several industrial processes, such as thermal spray coating deposition \cite{lavernia2010}, certain welding techniques \cite{david2003}, and metal additive manufacturing \cite{debroy2018}, operate under rapid solidification conditions. Rapid solidification leads to drastically altered microstructures through selection of metastable phases, solute trapping kinetics \cite{cahn1998,aziz1996,pinomaa2020MRS}, changes in solidification morphology  \cite{trivedi1994solidification}, and reduction of a microstructural feature size such as dendrite primary arm spacing \cite{trivedi1994solidification}. An overlooked, yet important, feature of these microstructures is the formation of various types of crystalline defects \cite{balluffi2005}. These include trapping of excess point defects \cite{zhang2017}, formation of high dislocation densities \cite{rukwied1971,wang2018,bertsch2020}, cavitation or microvoiding \cite{rukwied1971,pandey1986,michalcova2011} , high microstructural (type II-III) residual stresses \cite{chen2019,zhang2021multiscale}, and so-called lattice orientation gradients \cite{polonsky2020}. 
These crystalline defects critically affect the material's mechanical properties and performance. 
In this work, we focus on aluminium alloys. They are a light-weight class of materials relevant for e.g. transport industry, where rapid solidification conditions are known to significantly alter the mechanical properties of the material \cite{lavernia1992,lin2017,zhang2021multiscale,zhang2021evolution}.
In metal additive manufacturing of aluminium alloys, one of the key issues is controlling rapid solidification induced  shrinkage and the associated cracking \cite{ding2016,kotadia2021review}.

Solidification shrinkage leads to stresses in the forming solid, which induce plastic flow accompanied by formation of dislocations \cite{bertsch2020} and e.g. stacking faults \cite{voisin2021}. On continuum scale, the dislocations can be classified into geometrically necessary dislocations (GNDs) which are required to maintain lattice continuity when orientation variations are present, as well as statistically stored dislocations (SSDs) which occur through random trapping processes \cite{calcagnotto2010,muransky2019measurement}.   
SSDs have a net-zero Burger's vector, and thus do not affect the lattice curvature. It should be noted that the classification to GNDs and SSDs depends on spatial resolution. Even SSDs will cause lattice curvature in their immediate vicinity, if the atomistic crystalline lattice is fully resolved.

There are several experimental techniques to characterize dislocations. 
A view of individual dislocations can be obtained with transmission electron microscopy (TEM). The necessity to produce a thin foil sample is likely to disturb the original dislocation structure, and cause projection related inaccuracies \cite{bertsch2020}. Another technique, Electron channeling contrast imaging (ECCI), where the latter allows for one to characterize on significantly larger sample areas compared to TEM \cite{godec2020}. 
A commonly used technique to estimate dislocation densities is electron back-scatter diffraction (EBSD) imaging, where the local pixel-by-pixel crystalline orientation can be mapped out, and the misorientation between neighboring pixels can be used to estimate microstructure-scale lattice curvature, and through strain-gradient crystal plasticity theories, to dislocations (GNDs) necessary to accommodate this curvature \cite{calcagnotto2010}. 
%
Because the processes associated X-ray -- crystals interaction are simpler (mostly elastic) \cite{hammond2001}, the peak broadening in the resulting diffraction patterns can be more reliably used to estimate dislocation densities \cite{muransky2019measurement,bronkhorst2019} or e.g. lattice strains \cite{chen2019} in the illuminated volume at the high intensities produced in synchrotrons. Still, e.g. instrumental inaccuracies and sample preparation cause peak broadening, limiting the reliability of X-ray diffraction (XRD) based techniques to isolate the contribution of dislocation densities to the peak broadening \cite{hammond2001}.

Due to the limitations in the accuracy spatial and temporal resolution, simulation techniques provide valuable insight into the mechanisms behind formation of rapid solidification structures.
There have been numerous investigations into rapid solidification structures and their micromechanics of using molecular dynamics \cite{mahata2019_largescale_MD_solidification,mahata2019_solidification_defects_deformation}, phase field method \cite{keller2017,pinomaa2020_DTEM,pinomaa2020_acta}, crystal plasticity method \cite{bronkhorst2019,lindroos2021,francois2017}, and rapid solidification melt pool dynamics \cite{khairallah2016}. However, to our knowledge, formation of the aforementioned rapid solidification induced crystalline defects have not been investigated thoroughly with simulation approaches.

In this paper we consider pure and copper-alloyed aluminium, focusing on the formation of rapid solidification defects on multiple length scales, from MD to phase field crystal based amplitude expansion (PFC--AE) modeling, as well as coupled phase field -- crystal plasticity (PF--CP) simulations on the largest scales. Crystalline defect formation mechanisms are illustrated, and we compare the resulting average dislocation densities predicted by the considered modeling methods (MD, PFC--AE, PF--CP), and by past experiments. 

\section{Methods}

\begin{figure}[!h]
\centering\includegraphics[width=0.99\textwidth]{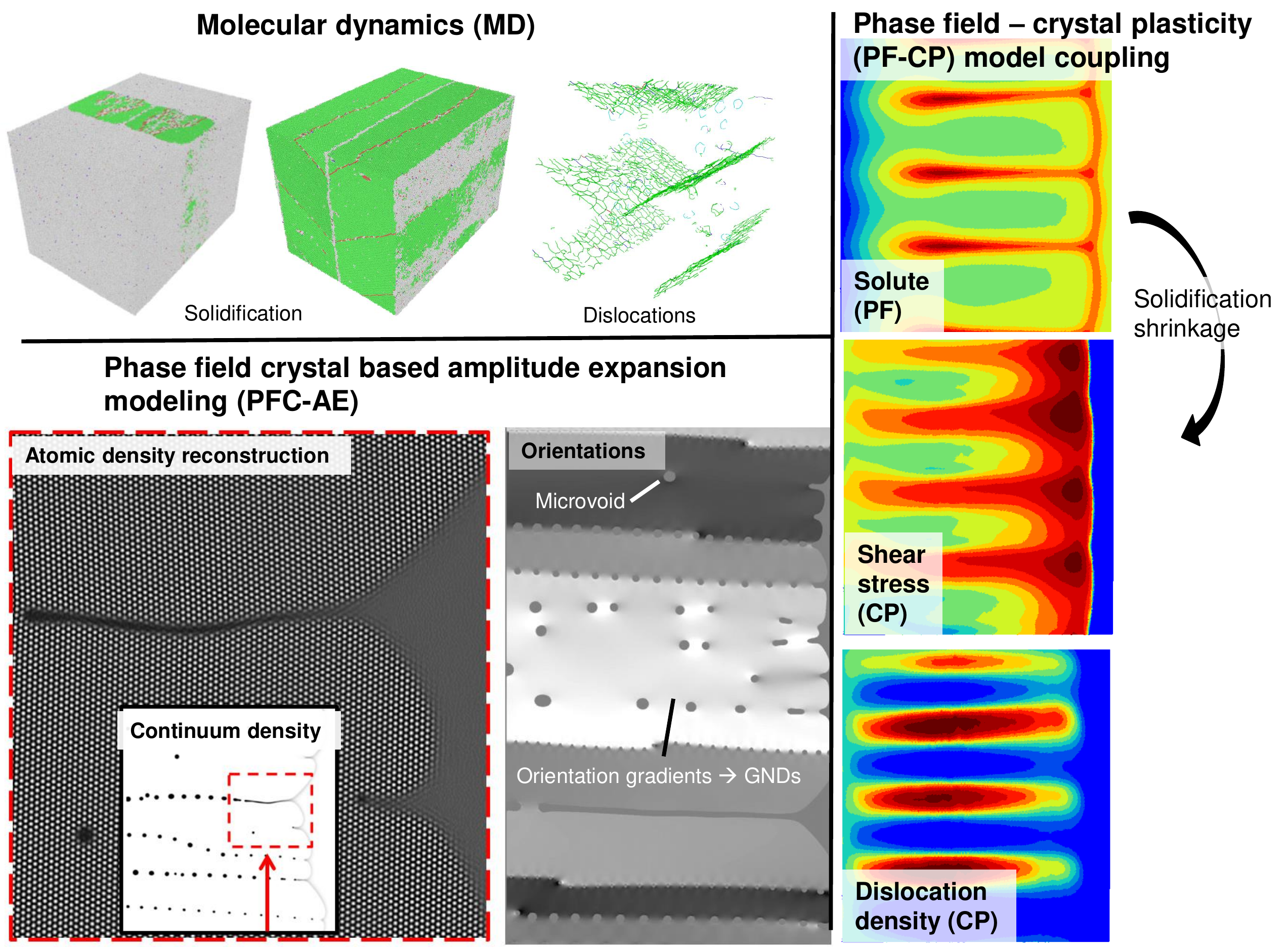}
\caption{Schematic of the multiscale modeling approach for describing crystalline defects forming in rapid solidification.}
\label{fig:flowchart}
\end{figure}

In this paper, we use a multiscale modeling framework presented in \autoref{fig:flowchart}, consisting of molecular dynamics (MD) simulations, phase field crystal based amplitude expansion (PFC--AE) simulations, and phase field -- crystal plasticity (PF--CP) simulations. 

\subsection{Molecular dynamics}
Molecular dynamics simulations of rapid polycrystalline solidification are carried out with LAMMPS \cite{plimpton1995}, where we use a bond-order potential for Al-Cu with an analytical form derived from quantum mechanics, accurate in the Al rich end of the binary system \cite{zhou2016}. When compared to experiments and ab-initio simulations, this bond-order potential has an excellent description of, for example, the pure aluminium  melting temperature, as well as the point defect characteristics and the characteristically high stacking fault energy of aluminium. The phase diagram resulting from this bond-order potential is presented elsewhere \cite{haapalehto2021}.

\begin{figure}[!h]
\centering\includegraphics[width=0.95\textwidth]{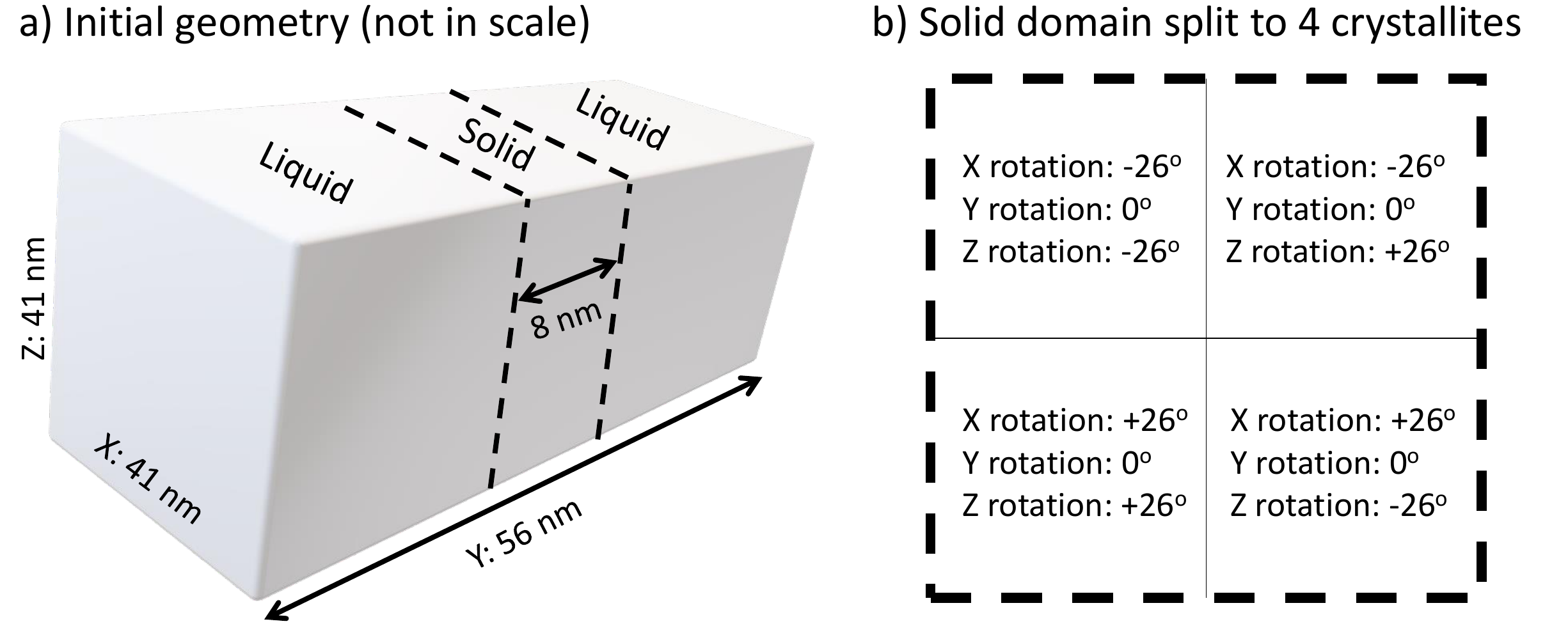}
\caption{a) Initial geometry for the molecular dynamics simulations, and b) the four rectangular crystallites initialized in the solid domain.}
\label{fig:MD_geometry}
\end{figure}

The free solidification conditions are initialized to contain a co-existing solid and liquid phases as depicted in \autoref{fig:MD_geometry}, following Yang et al. \cite{yang2011}, adjusted to polycrystalline solid domain of four misoriented crystallites. For the initialization, first a solid domain of pure aluminium FCC is generated in a cuboid geometry as shown in \autoref{fig:MD_geometry}a, where the solid is split into four misoriented grains (\autoref{fig:MD_geometry}b). Afterwards, randomly selected aluminium atoms are replaced by copper atoms such that the compositions of the solid and liquid regions are approximately equal to the solidus and liquidus values corresponding to the equilibration temperature; the MD-based phase diagram is presented elsewhere \cite{haapalehto2021}. Then, atoms in the solid region are fixed, while the rest of the system is melted at a relatively high temperature in an $\mathrm{NP_YAT}$ ensemble, where $Y$ is the solidification direction. In this ensemble, the total number of particles (N) and cross-sectional area (A) are kept constant, while the average temperature (T) and longitudinal pressure ($\mathrm P_\mathrm{Y}$) are controlled by a Nosé-Hoover thermostat-barostat.
As the four crystallites (\autoref{fig:MD_geometry}) contain grain boundaries with high curvatures, we do not perform any relaxation at this point in order to preserve the grain boundary structure. 

The box dimensions are 41 nm $\times$ 41 nm $\times$ 56 nm, where the longest dimension is the direction where solidification progresses, and a solid region was initialized in the middle. Periodic boundary conditions are imposed in all directions. The solid region was segmented into four domains where the lattice orientations were rotated as indicated in \autoref{fig:MD_geometry} b.

The solidification is started by setting the solute concentration in the liquid region to a chosen value and setting the thermostat temperature to the chosen undercooling. 
The solidification simulations are performed in a $\mathrm{NP_YAT}$ ensemble using a single global Nosé-Hoover thermostat. It should be noted that the latent heat release leads to local temperature rise above the thermostat target temperature. 
A detailed analysis of single crystal Al-Cu rapid solidification kinetics, point defects, and chemical ordering is presented in a separate article \cite{haapalehto2021}.

\subsection{Phase field crystal based amplitude expansion simulations}
Phase field crystal based amplitude expansion (PFC-AE) simulations were conducted for rapid solidification of pure aluminium.
In contrast to traditional phase field methods\cite{boettinger2002,pinomaa2020_DTEM}, PFC is capable of self-consistently retaining lattice orientation and other atomic information while evolving microstructure dynamically on diffusive time scales \cite{elder2007}. 

In this work, we use a PFC model based on the one proposed by Kocher et al. \cite{kocher2015}, which accounts for both liquid and vapor disordered phases as well as an ordered solid phase in a system consisting of a single atomic species. The existence of a vapor phase in the model is crucial under the assumption that extreme interface velocities in a pure material system would be capable of trapping microvoids \cite{rukwied1971,pandey1986,michalcova2011}, in a manner analogous to solute trapping in alloy systems \cite{pinomaa2020MRS}.

The PFC model exhibits a thermodynamic phase diagram approaching that of pure aluminum in the vicinity of the triple point \cite{jreidini2022}.
The reference density was set near the solid coexistence line at a temperature appropriate for deep quenches. The model's phase diagram exhibits a sharp deviation from pure aluminum's at densities below the critical point, due to the expanded nature of this model's free energy functional \cite{elder2007}. Further work is underway to expand PFC approaches to more accurately span the range of densities accessible in pure materials' phase diagrams \cite{kocher2019}. As the present work focuses on solidification, good agreement with the solidus in the vicinity of the triple point was determined to be sufficient.

Although PFC models operate in diffusive time scales, they are still limited in spatial scale as individual atomic density peaks need to be resolved. To achieve larger simulation system scales, the PFC model is coarse-grained into a complex order parameter phase field model, where individual atomic peaks are integrated out, through the use of the amplitude expansion method \cite{oforiopoku2013,athreya2006}. 

This amplitude field model was simulated in a co-moving reference frame where circular randomly oriented grains were initialized with an up to 10 degree misorientation with respect to the solidification direction.
Periodic boundaries were used for system edges perpendicular to the solidifying front. 
More details about rapid solidification phase field crystal simulations are presented in another publication \cite{jreidini2021}.

\subsection{Phase field method with crystal plasticity}

\subsubsection{Phase field model for rapid solidification}
On the largest scale (several micrometers), rapid solidification is simulated with phase field method. Solute trapping kinetics according to continuous growth model \cite{aziz1996} is implemented through a modified anti-trapping current, as described in \cite{pinomaa2019_acta}, in a multi-order parameter model \cite{ofori2010}, and applied to thin film rapid solidification conditions \cite{pinomaa2020_DTEM}. 
For the phase field model presented in \cite{pinomaa2020_DTEM}, the model parameters are summarized in \autoref{table:phaseField_parameters}. For the temperature field, a frozen gradient approximation is imposed, representative for rapid solidification conditions. 
More details about this phase field model implementation are presented in Ref. \cite{pinomaa2020_DTEM} with slightly different model parameters corresponding to a dilute Al-Cu alloy, and in Ref. \cite{lindroos2021} for 316L stainless steel.
The simulations are carried out with Z-set software \cite{ammar2009,de2016}, where the information about the order parameter and concentration fields are forwarded to crystal plasticity simulations, whose the methodological details are summarized in what follows.

\begin{table}
\caption{ Phase field model parameters for Al-4-5at\%Cu alloy.}
\begin{tabular}{lcc}
\hline
Equilibrium partition coefficient $k_e$  & 0.103$^a$  
\\
Equilibrium liquidus slope $m_l^e$   & -0.579 K/at\%$^a$
\\
Alloy concentration $c_l^o$    & 4.5 at\%
\\
Solid-liquid interface energy $\sigma_{SL}$ & 0.127 J/m$^2$ $^b$
\\
Gibbs-Thomson coefficient $\Gamma$   & 1.098$\times$ 10$^{-7}$ K m  $^b$
\\
Solutal capillary length $d_0$   & 5.1 nm $^b$
\\
Interaction parameter $\omega$   & 2.0 
\\
Liquid diffusion coefficient $D_L$ & 3.0 $\times $10$^{-9}$ m$^2$/s 
\\
Solid diffusion coefficient $D_s$  & 0 m$^2$/s
\\
Kinetic coefficient $\beta_0$           & 0.04466 s/m 
\\
Capillary anisotropy strength $\epsilon_c$       & 0.01 $^c$
\\
Kinetic anisotropy strength $\epsilon_k$  & 0.13 $^d$
\\
Solute trapping velocity $V_D^{PF}$ & 4.9 m/s
\\
Computational interface width $W_0$ & 6 nm 
\\
Trapping parameter $A$& 0.125
\\
Element size $dx$ & 1.0$W_0$
\\
\hline
\end{tabular}
\\
$^a$: Estimated using Thermo-Calc's TCAL6 database \cite{andersson2002} \\
$^b$: From molecular dynamics simulations in Ref. \cite{haapalehto2021}. \\
$^c$: From experiments in Ref. \cite{liu2001} \\
$^d$: From molecular dynamics simulations in Ref. \cite{bragard2002}
\label{table:phaseField_parameters} 
\end{table}

\subsubsection{Crystal plasticity model}
To investigate defect formation in rapid solidification on microstructural scales resulting from solidification shrinkage, a crystal plasticity model is presented closely following the model by Monnet et al. \cite{monnet2019}. We impose a small deformation framework, where the total strain ($\underline{\epsilon}$) is additively decomposed to elastic ($\underline{\epsilon}^E$), plastic ($\underline{\epsilon}^P$), solidification shrinkage ($\underline{\epsilon}^S$), and thermal ($\underline{\epsilon}^T$), contributions, respectively:
\begin{equation}
    \underline{\epsilon} = \underline{\epsilon}^E + \underline{\epsilon}^P + \underline{\epsilon}^S + \underline{\epsilon}^T.
\end{equation}
Dislocation slip is assumed to be the dominant deformation mechanism. 
As these simulations focus on the solidification, the system temperatures are close to the alloy solidus temperature, and solid-state thermal contraction (due to the solid's thermal expansion effect) is insufficient to produce notable thermal thermal strains within the simulation box. 

The strain rate due to solidification shrinkage is given by 
\begin{align}
    \underline{\dot{\epsilon}}^S &= \underline{\epsilon}_0 \frac{1}{2} \left( \dot{h} + 1 \right) \underline{I},
\end{align}
where $\underline{\epsilon}_0$ is the axial dilation (shrinkage) strain adjusted to produce approximately 7 \% volumetric shrinkage from liquid to solid, $h=h( \lbrace \phi_i \rbrace)$ indicates the solid fraction based on the order parameters drawn from the phase field model:
\begin{align}
    h(\lbrace \phi_i \rbrace) &= N_{orient} - 1 + \sum_{i=1}^{N_{orient}} \phi_i,
    \label{eq:CP_h_solidLiquidIndicator}
\end{align}
where $h \rightarrow 1$ in the solid and $h \rightarrow -1$ in the liquid, and in-between values in the solid-liquid interface. 
The net elasticity tensor ($\Lambda$) is interpolated between liquid and orientation of each solid phase as 
\begin{align}
    \Lambda = \frac{1-h}{2} \Lambda_{L} + \sum_{i=1}^{N_{orient}} \frac{1+\phi_i}{2} \Lambda_i,
\end{align}
where $\Lambda_{L}$ is the liquid elasticity tensor, and $\Lambda_i$ is the elasticity tensor of a crystal corresponding to order parameter $\phi_i$ solid region.

Plastic strain rate is given by the sum of twelve $\{111\}<110>$ slip systems in FCC aluminium.
The plastic strain rate is thus defined as 
\begin{equation}
 \underline{\dot{\epsilon}}^P =\sum_{s=1}^{N_s} \dot{\gamma}^s \underline{N}^s 
\end{equation}
where $N_s$ is the number of slip systems, and $\underline{N}^s$ , and viscoplastic flow rule yields a shear rate for slip system $s$ as
\begin{equation}
    \dot{\gamma}^s = \frac{1+h(\lbrace \phi_i \rbrace)}{2} \left\langle \frac{|\tau^s| - \tau_c^s}{K_{\tau}} \right\rangle^N sign(\tau^s),
    \label{eq:slip_rate}
\end{equation}
where $\tau^s$ is the resolved shear stress and $\tau_c$ is the effect slip resistance of a slip system $s$, and the solid fraction $h(\lbrace\phi_i\rbrace)$ was used as a prefactor in \autoref{eq:slip_rate} to ensure that plastic deformation does not occur in the liquid. Viscous parameter $K_{\tau}$ is chosen to be equal to the solid solutions strength of the material, and parameter $N$ defines the strain rate dependence. Critical resolved shear stress is defined as
\begin{equation}
    \tau_c^s = \tau_{0}^{SS} + \tau_{HP} + \mu b^s \sqrt{\sum_j a_{sj} \rho^j},
    \label{eq:tau_c}
\end{equation}
where $\tau_{0}^{SS}$ is the dislocation-free shear stress resistance for large grains with solid solution strengthening effects included. The Hall-Petch-like size effect can be described with $\tau_{HP}$ in \autoref{eq:tau_c}; for simplicity, and due to the lack of available data, we set the size effect strengthening to zero in the actual simulations: $\tau_{HP} = 0$. However, when the single crystal parameters were fitted with experimental stress-strain data (fitting procedure described further below), the HP effect is included to match the initial yield behavior of the material as follows:
\begin{equation}
    \tau_{HP} = \frac{\mu(T)}{\mu_{ref}} \frac{K_{HP}}{\sqrt{d_{grain}}},
    \label{eq:CP_tau_HP}
\end{equation}
where $\mu(T)$ is the temperature-dependent shear modulus, $\mu_{ref}$ is the shear modulus in room temperature, $K_{HP}$ is Hall-Petch coefficient, and $d_{grain}$ is the effective grain size.

The last term on the right-hand side of \autoref{eq:tau_c} leads to dislocation-induced forest hardening, where $\mu$ is the shear modulus, $b^s$ is the length of Burgers vector for slip system $s$, $a_{sj}^{\tau}$ interaction matrix between slip systems $s$ and $j$, and $\rho^j$ dislocation density of slip system $j$. 
Once deformation and plasticity due to solidification takes place, dislocation density can increase rapidly. 


In this paper, we assume that solid solution strengthening effect in Al-Cu due to copper is negligible, and therefore the resistance coming from this source was set to zero in the simulations. However, the solid solution term can have significant effect on slip resistance in other materials and the key aspect then is to retrieve the local concentration field from phase field simulations, as presented elsewhere for 316L steel PF--CP simulations \cite{lindroos2021}.



The evolution of statistically stored dislocation density corresponding to slip system $s$ is governed by 
\begin{equation}
    \dot{\rho}^s = \frac{|\dot{\gamma}^s|}{b^s} \left[\frac{1}{d_{grain}} + \frac{1}{K_{forest}} \sqrt{\sum_j^{forest} a_{sj} \rho_j}  + \frac{1}{K_{coplan}} \sqrt{\sum_{j}^{coplan} a_{sj} \rho_j} - G_c \rho^s \right],
    \label{eq:dislocation_evolution}
\end{equation}
where $d_{grain}$ is the mean grain size (used in polycrystal parameter fitting),  $K_{forest}$ and $K_{coplan}$ define the hardening effect from dislocations. They may be interpreted to describe the average number of obstacles passed by a dislocation before immobilizing. Dislocation annihilation is controlled with a parameter $G_c$.

The CP model was parametrized using data in Ref. \cite{rudra2020}, where elevated temperature polycrystalline stress-strain data of aluminium alloy in series 5083 at 673 K and a strain rate of 10 s$^{-1}$. We chose this alloy as it should have negligible strengthening effects due to precipitates, which are neglected in this work.  For the fitting procedure, we chose viscous parameter values $K = 8$ and $N=15$ in \autoref{eq:slip_rate}, whereas for the PF--CP simulations presented later, we used $K=1$ and $N=15$. For the fitting, we activated the Hall-Petch strengthening in \autoref{eq:CP_tau_HP} with $K_{HP} = 0.15$ and $d_{grain}= 180$ $\mu$m; in the PF--CP simulations we set $\tau_{HP} = 0$. We initialized a simple equiaxed polycrystalline structure, and using the stress-strain data in Ref. \cite{rudra2020}, we set the dislocation hardening parameters to $K_{forest}=50$ and $K_{coplan}=150$. The used parameter values are summarized in \autoref{table:CP_parameters}.

\begin{table}
\centering
\caption{Crystal plasticity model parameters for Al-4-5at\%Cu alloy.}
\begin{tabular}{lcc}
\hline
Young's modulus $E_S$ & 31.5 MPa $^a$ 
\\
Poisson's ratio (solid) $\nu^S$    & 0.27 $^b$
\\
Shear modulus (solid) $\mu^S$   & $E / (2(1+\nu^S))$  
\\
Young's modulus (liquid) $E^L$  & 0.1 GPa
\\
Poisson's ratio (liquid) $\nu^L$    & 0.49
\\
Length of Burgers vector $b^s$   & 0.296 nm
\\
Forest obstacle parameter $K_{forest}$   & 50.0
\\
Coplanar obstacle parameter $K_{coplan}$   & 150.0
\\
Annihilation distance $G_c$ & 2.96 nm
\\
Viscous parameter for slip $K_{\tau}$ &  1.0 
\\
Viscoplastic strain rate exponent $N$ & 15.0
\\
Solidification strain components $\epsilon^s_{1,2,3}$ & -0.0228 $^d$
\\
Interaction matrix $a_{sj}$ & See Table 1 in Ref. \cite{monnet2019}                       \\
\hline
\end{tabular}
\\
$^a$: Interpolated value at 674 K for Al 6061 alloy from Ref. \cite{summers2015}. \\
$^b$: Interpolated value at 674 K for FCC 316L steel from Ref. \cite{kusnick2013}. \\
$^c$: Interpolated value at 674 K for Al 5083 alloy from Ref. \cite{maljaars2005}. \\
$^d$: Adjusted to produce a 7\% volume decrease from liquid to solid.
\label{table:CP_parameters}
\end{table}

The PF--CP simulations are performed on Z-set software \cite{ammar2009,de2016}, where more detailed simulations for a case of 316L steel are found elsewhere \cite{lindroos2021}.

\section{Results}

\subsection{Molecular dynamics (MD)}
\begin{figure*}
\centering\includegraphics[width=0.99\textwidth]{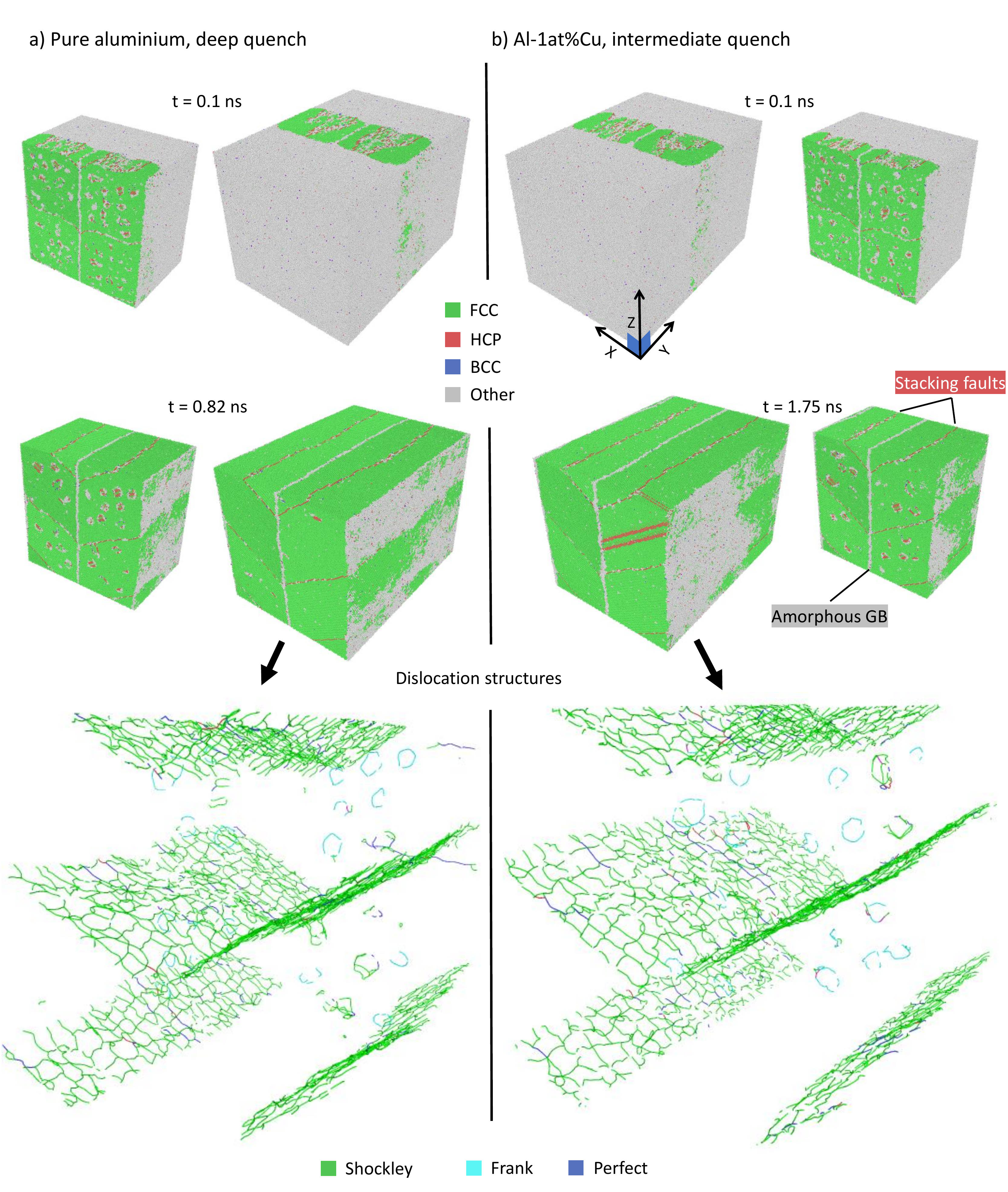}
\caption{Atomistic simulations of polycrystalline aluminium rapid solidification, showing the time evolution and the dislocation structure of the solidified structures, where a) is pure aluminium (interface temperature $\approx$ 897 K, soldification velocity $\approx$ 56 m/s), and b) is Al--1at\%Cu (interface temperature $\approx$ 898 K, solidification velocity $\approx$ 25 m/s).}
\label{fig:MD}
\end{figure*}
In the MD simulations, a simulation box with liquid in the middle, and the resulting rapid solidification at early time (0.1 ns) and fully solidified structures are shown in \autoref{fig:MD} for pure aluminium and Al--1at\%Cu. The full simulation geometry is shown side by side with the geometry clipped in half in the solidification ($Y$) direction. The different phases are identified with Ovito's Polyhedral Template Matching (PTM) algorithm \cite{ovito}, where the gray color (labeled "Other") corresponds to an unidentified coordination number, which is predominantly liquid or amorphous phase. The grain boundaries with normal in $X$ direction remain more or less similar throughout the solidification, maintaining the disordered structure. In contrast, the grain boundaries in $Z$ direction quickly form stacking faults, seen as tilted grain boundary with HCP (red) interface layers. Stacking faults have been observed in past Al-Cu rapid solidification MD simulations \cite{mahata2019_largescale_MD_solidification}.

\autoref{fig:MD} also shows the dislocation structures as identified with Ovito's Dislocation analysis tool, DXA \cite{stukowski2012}. The single layer stacking faults contain Shockley partial dislocations, shown as wedge-shaped red layers, which continue in $Z$ and $X$ direction due to the periodic boundary conditions. In addition, Frank partials and perfect dislocations are identified in the interiors of each four crystallite. Also, a small number of stair-rod and Hirth type dislocations are identified. The resulting dislocation densities are summarized in \autoref{table:dislocations}, and are compared to other simulation techniques and literature values in \autoref{section:discussion}. 

Rapid solidification kinetics of single crystal pure aluminium and aluminium-copper alloy is presented in another paper \cite{haapalehto2021}, where we find high interstitial and vacancy concentrations, but no dislocations. This suggests that the generation of dislocations is intimately tied to interaction between misoriented grains in rapid solidification of atomistically small systems. 
The role of misorientations, orientation gradients, and GNDs, are further investigated with PFC--AE simulations in what follows. 


\subsection{Phase field crystal method based amplitude expansion (PFC--AE) simulations}


In the PFC--AE simulations, we considered cases of rapid solidification and an intermediate solidification velocity, for pure aluminium. For the rapid solidification case, we applied a frozen temperature gradient $G=\SI{5e7}{\kelvin\per\meter}$ with an average system temperature set to $T=\SI{805}{\kelvin}$ initially, a significant quench below pure aluminum's melting point of approximately $\SI{933}{\kelvin}$, and the average density was chosen close to the solid's density; we refer to this case as a "large quench". A second case, leading to lower solidification velocities--referred to as "intermediate quench"---, was carried out, where a quench temperature $T=\SI{875}{\kelvin}$ was chosen in the limit of no  temperature gradient, i.e., $G=0$. The lattice orientations in the simulated solid grains were calculated from the atomic displacement field extracted from the complex amplitude fields following Ref. \cite{heinonen2016}.

\begin{figure*}
\centering\includegraphics[width=0.99\textwidth]{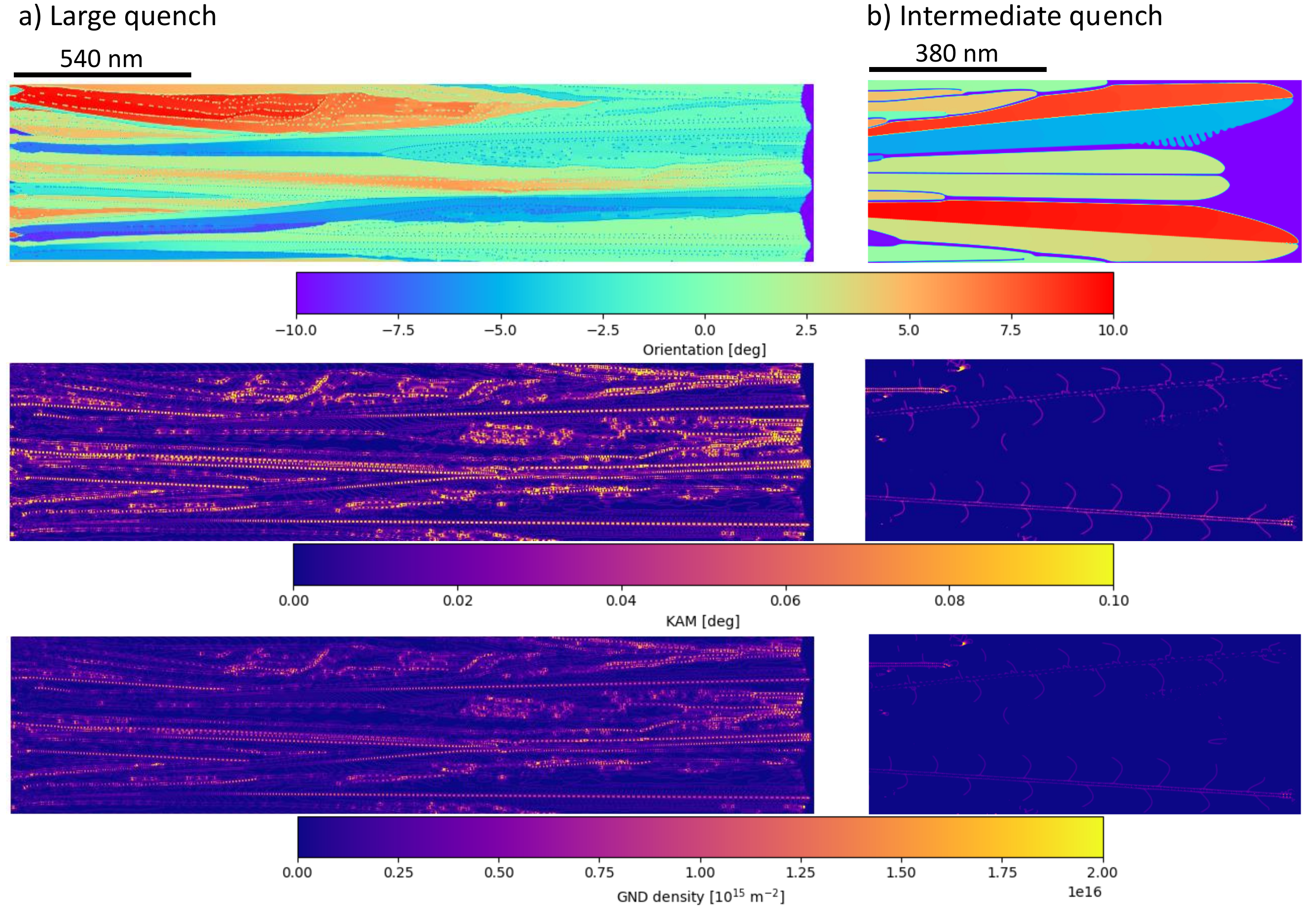}
\caption{Phase field crystal simulations of pure Al rapid solidification, with the kernel average misorientations (KAM), as well as the geometrically necessary dislocation (GND) densities, where a) corresponds to large quench (large solidification velocity), and b) to an intermediate quench and velocity.}
\label{fig:PFC}
\end{figure*}

\autoref{fig:PFC} (top row) shows the lattice orientations in a post-solidified simulated sample, reconstructed from snapshots of the co-moving simulation frame. The large quench case leads to smaller microstructural features and the highly misoriented grains (close to orientation +10$^o$ and -10$^o$) tend to be dominated by grains that have an orientation closer to the solidification axis (0$^o$). Moreover, \autoref{fig:PFC}a has visible orientation gradients, where it can be seen that the orientation distribution tends towards the solidification direction as the solidification progresses. This rapid solidification induced formation of low angle grain boundaries has been seen earlier for melt-spun aluminium alloys \cite{lin2017}, as well as in additively manufactured 316L steel \cite{polonsky2020}. In contrast, the intermediate quench case leads to clearly separated cellular tips where the orientation does not vary notably. The orientation gradients are compared to thin film rapid solidification experiments elsewhere \cite{jreidini2021}.

The orientation gradients were quantified in terms of kernel average misorientation (KAM), shown in the middle row of \autoref{fig:PFC}. To calculate the KAM, the raw simulation data was first downscaled by subdividing the simulation data into blocks, each containing 8x8 grid points, where the local orientation was calculated as the average of each block. For the downscaled blocks, KAM was then evaluated by summing the absolute value of local disorientations up to fourth nearest neighbors, excluding disorientations larger than 2 degrees following \cite{calcagnotto2010}, as well as pixels where the density was close to the liquid value, which occurred at high-angle grain boundaries, the liquid domain, as well as small liquid or vapor pockets (discussed further below). As the laplacian of the density field has a large magnitude at the liquid or vapor interfaces, we chose an appropriate cut-off value for the density field laplacian to mask out interfacial pixels from the KAM evaluation. The resulting KAM distributions are shown in \autoref{fig:PFC} (second raw), illuminating large KAM hot spots throughout the solidified system for the large quench case, whereas only a few KAM hot spots are visible in the intermediate quench case. 

Maintaining strain compatibility across a polycrystalline structure with anisotropic mechanical properties occurs by intragrain lattice curvature, which necessitates arrays of appropriate dislocations with net non-zero Burger's vector. These dislocations are called geometrically necessary dislocations (GND). Through energy minimization, these dislocations tend to arrange themselves into cells, dividing a grain into subgrains, which can also be called geometrically necessary boundaries \cite{muransky2019measurement}. These  subgrains or "inner dendrites", corresponding to small rotations caused by thermal stresses, are visible also in FCC 316L steel in rapid solidification through additive manufacturing \cite{godec2020}.

Following Ref. \cite{calcagnotto2010}, the KAM distribution ($\theta_{KAM}$) was used to determine the GND density distribution:
\begin{align}
    \rho_{GND} = 2\frac{\theta_{KAM}}{r_{kernel} b},
\end{align}
where $r_{kernel}$ is the kernel radius and $b$ is the Burgers vector length. As the KAM was evaluated using fourth nearest neighbors, $r_{kernel} \approx 2 a_{Al}  $, where $a_{Al} \approx 0.4196$ nm is pure aluminium FCC lattice constant near its melting point \cite{shin2017}. Assuming dislocations due to $\lbrace 111 \rbrace \langle 110 \rangle$ slip, the Burgers vector magnitude is $b = a_{Al}/\sqrt{2}$. The resulting GND distribution is shown in \autoref{fig:PFC}, illustrating that the large quench case leads to subdivision into subgrains through high KAM and GND densities, whereas for the intermediate quench case, GND and KAM levels are low and division into subgrains is negligible.
The average GND densities are listed in \autoref{table:dislocations} for the large and intermediate quench cases, and compared to other simulation techniques and literature later in \autoref{section:discussion}.

\begin{figure*}
\centering\includegraphics[width=0.99\textwidth]{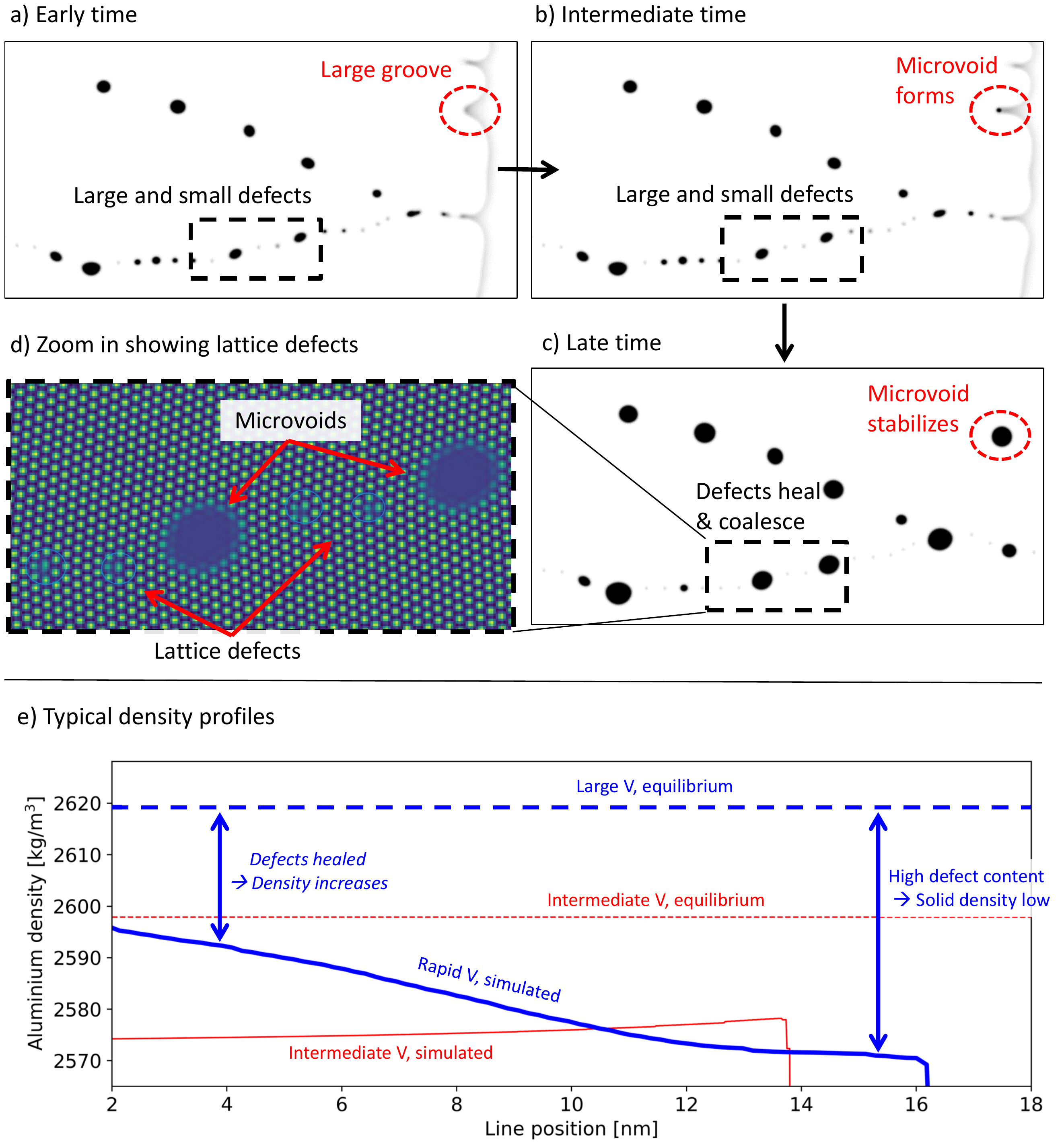}
\caption{Phase field crystal based amplitude expansion simulations of rapid solidification: a-c) show time evolution of the density field, illustrating defect pocket formation, merging, and healing; d) shows a zoom-in of the reconstructed atomic lattice, showing stabilization of lattice defects and microvoids; e) shows typical density line profiles in the solid side close to the liquid.}
\label{fig:PFC_defects}
\end{figure*}

The slower solidification conditions (\autoref{fig:PFC}a) result in relatively smooth solid-liquid interfaces with few defects trapped in the solid. In contrast, the higher driving force case (\autoref{fig:PFC}b) displays rougher interfaces where the interface grooves trap disordered pockets of fluid that can later become stable defects. The trapping of defects from interface grooves, and the resulting defect stabilization, is exemplified in \autoref{fig:PFC_defects}a--c (red dashed circles). This defect formation mechanism is consistent with molecular dynamics simulations by He et al. \cite{he2019}, where a notable solid-liquid interface roughness--albeit on smaller length scales compared to our simulations--facilitates the formation of crystalline defects, by allowing for deep solid-liquid interface grooves to `fold in' and trap non-crystalline domains as stable defects. 
There is indirect experimental evidence of rapid solidification induced microvoid formation, or cavitation, in  melt-spun aluminium alloys \cite{michalcova2011}, a powder metallurgically solidified Al-Li alloy \cite{pandey1986}, and cast high purity copper \cite{rukwied1971}. It should be noted that an alternative formation mechanism for these experimentally observed microvoids is that trace amounts of gases, from the processing atmosphere, can dissolve into the metal melt and form microscopic bubbles upon solidification.

It is also interesting to analyze the healing of crystalline defects after the defect pockets have formed. Smaller defect pockets heal into perfect crystals, or coalesce with existing microvoids, as shown in \autoref{fig:PFC_defects}a--c (dashed rectangles). The reconstructed atomic lattice is presented in  \autoref{fig:PFC_defects}d, illustrating two stable microvoids and smaller lattice defects.

Density profile lines were extracted in the vicinity of the solid-liquid interface, as shown in \autoref{fig:PFC_defects}e, for both the large quench (blue solid line) and intermediate quench (red solid line) cases both according to the PFC--AE simulations. The corresponding equilibrium values are shown as dashed horizontal lines. It can be seen that the larger quench leads to a relatively large deviation from the material equilibrium density due to a large number of defect pockets that form during rapid solidification. As the defect pockets heal, the solid density starts to approach the equilibrium density. In contrast, the intermediate quench does not form notable defect pockets, and therefore the density remains approximately constant after solidification. The abrupt drop in the PFC--AE simulation density profiles (red and blue solid lines in \autoref{fig:PFC_defects}e) correspond to the occurence of the solid-liquid interface.
More details on the orientation gradients using this PFC-AE model are presented elsewhere \cite{jreidini2021}.

\subsection{Phase field -- crystal plasticity (PF--CP) simulations}

\subsubsection{Phase field simulations}

The phase field simulations are carried out in a rectangle of size 480 $\mu$m $\times$  960 $\mu$m, with zero flux boundary conditions on all edges for the order parameter and concentration fields. A frozen gradient is applied with a pulling speed $V=$0.03 m/s, and either $G =$ 10$^6$ K/m or 2$\times$10$^6$ K/m; also, we initialized either a single crystal three distinct orientations. The single crystal cases are initialized as sinusoidal solid fronts, and the three distinct crystals are initialized as circles. 

\begin{figure}[!h]
\centering\includegraphics[width=0.99\textwidth]{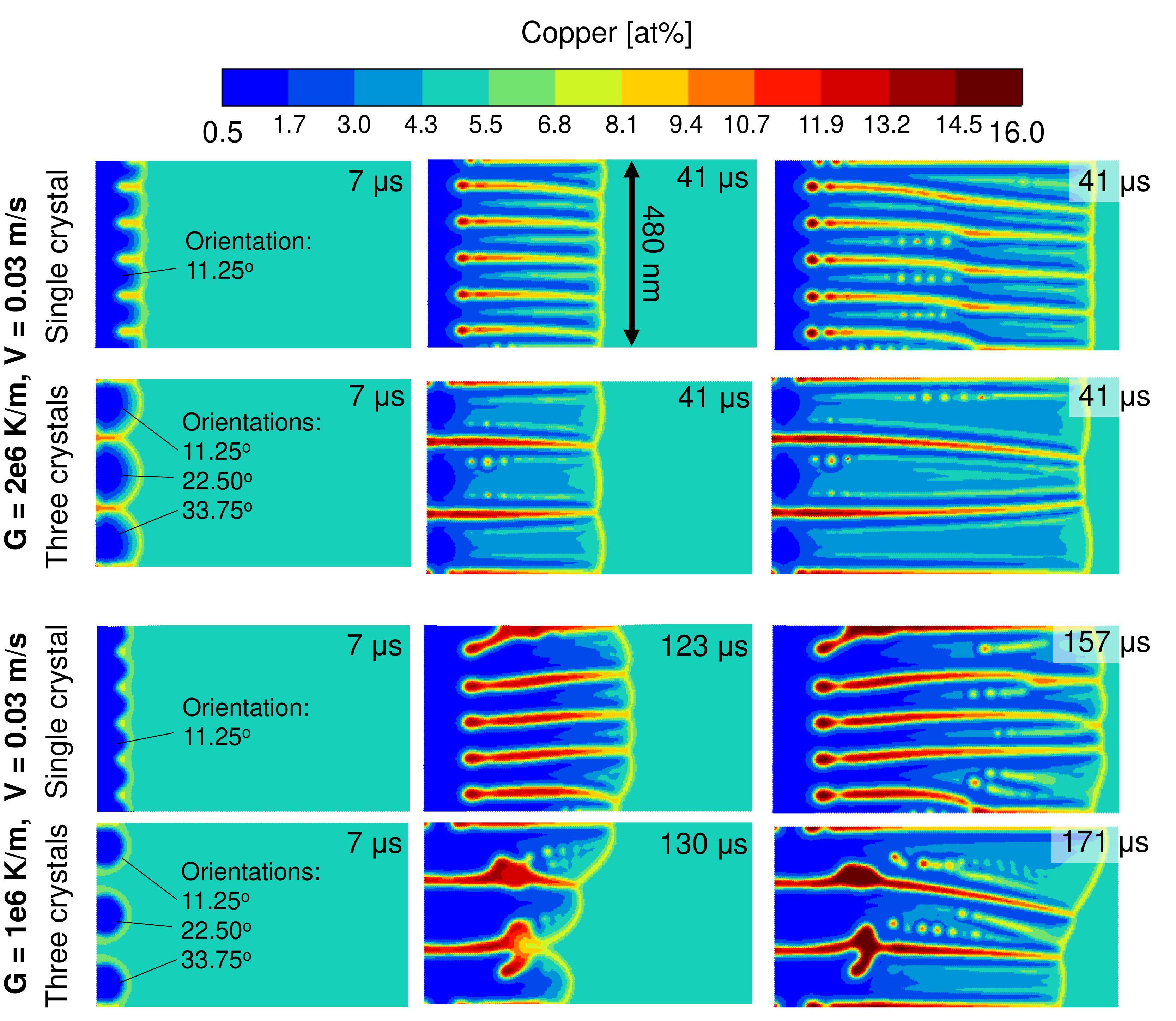}
\caption{Phase field simulations of rapid solidification for Al--4.5at\%Cu with two different thermal gradients $G$, and with either a single crystal or three crystals initialized.}
\label{fig:PF_CP_Cs_allCases_evolution}
\end{figure}

The time evolution of each four case is shown in  \autoref{fig:PF_CP_Cs_allCases_evolution} in terms of the copper concentration. 
The single-crystal simulations approach a quasi-steady state cellular growth mode towards the end of the system far-end boundary, with the smaller $G$ resulting in a larger cell spacing as expected. The three crystal simulations lead competitive growth of misoriented grains, where the larger $G$ leads to relatively straight grain boundary trajectories as the capillary anisotropy becomes less important at larger undercoolings \cite{bragard2002}.  Overall, the smaller $G$ leads to more heterogeneity in the solidification microstructures, visible as stronger microsegregation. This heterogeneity is reflected in the microstructural residual stress distribution through results from the solidification shrinkage, analyzed further below.

\subsubsection{Crystal plasticity coupling}
The crystal plasticity simulations were carried out by taking the  order parameter fields from the phase field simulations shown above, and coupling their dynamics to the crystal plasticity model. Specifically, the order parameter time evolution resulted in  solidification shrinkage induced stresses, which lead to to plastic deformation and generation of statistically stored dislocations (SSDs). The left and bottom boundary nodes were fixed in $X$ and $Y$ directions, respectively. The top and right boundary nodes were allowed to move, while staying on a straight line through the multi-point constraint technique. 
The initial dislocation density was set uniformly to a small non-zero value, 10$^7$ m$^{-2}$. 

\begin{figure}[!h]
\centering\includegraphics[width=0.99\textwidth]{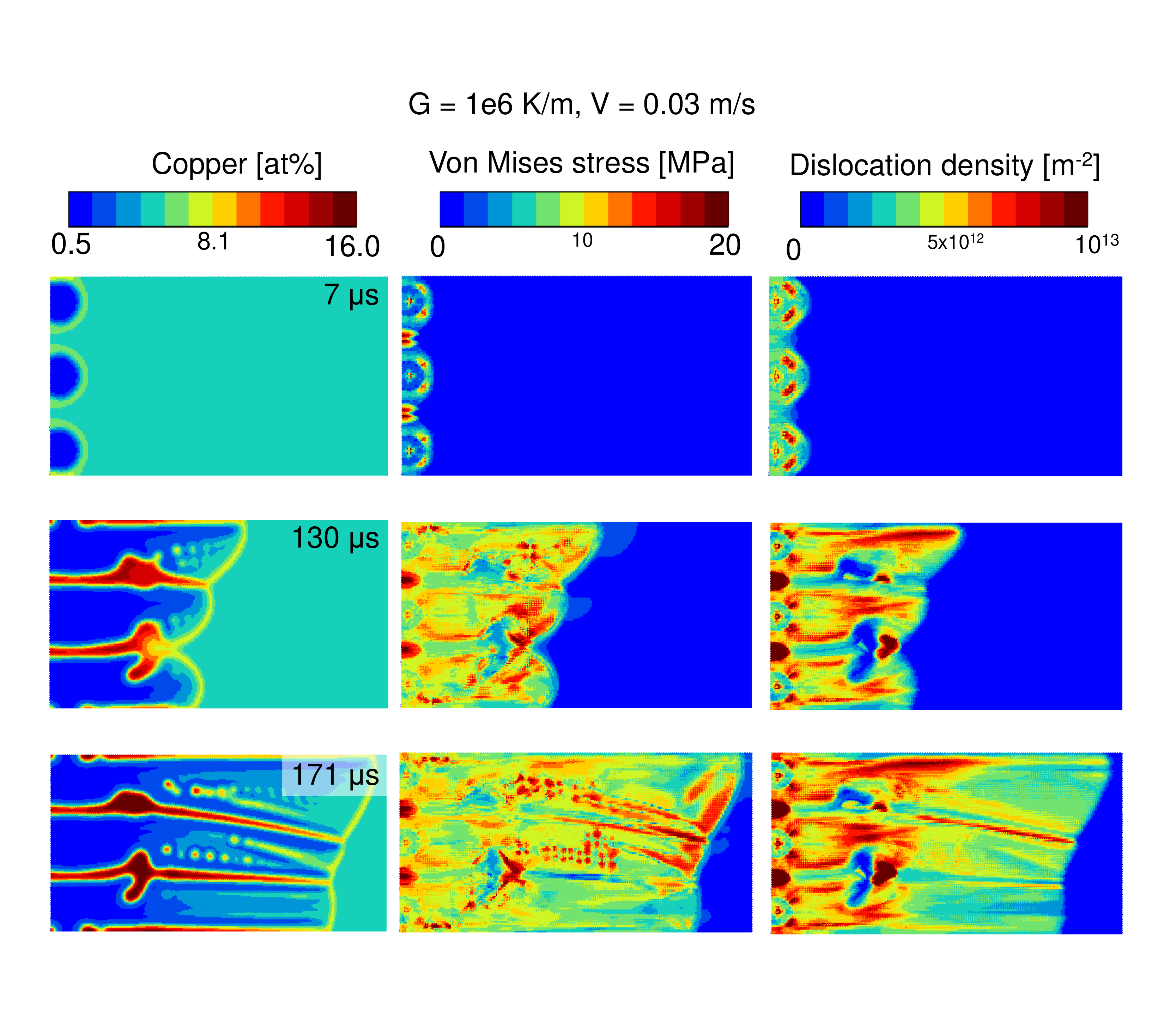}
\caption{Phase field -- crystal plasticity coupled simulation of Al--4.5at\%Cu rapid solidification, showing the copper concentration, solidification shrinkage induced von Mises stresses, and the densities of statistically stored dislocations.}
\label{fig:PF_CP_Cs_mises_rhos_evolution}
\end{figure}

The evolution of concentration (from phase field simulations), von Mises stresses, and dislocation densities are shown for the $G$ = 10$^6$ K/m case with three crystals in \autoref{fig:PF_CP_Cs_mises_rhos_evolution}. The solidification shrinkage generates von Mises hot spots that follow the heterogeneity in the microsegregation, with intercellular regions having higher SSD densities. 

\begin{figure}[!h]
\centering\includegraphics[width=0.99\textwidth]{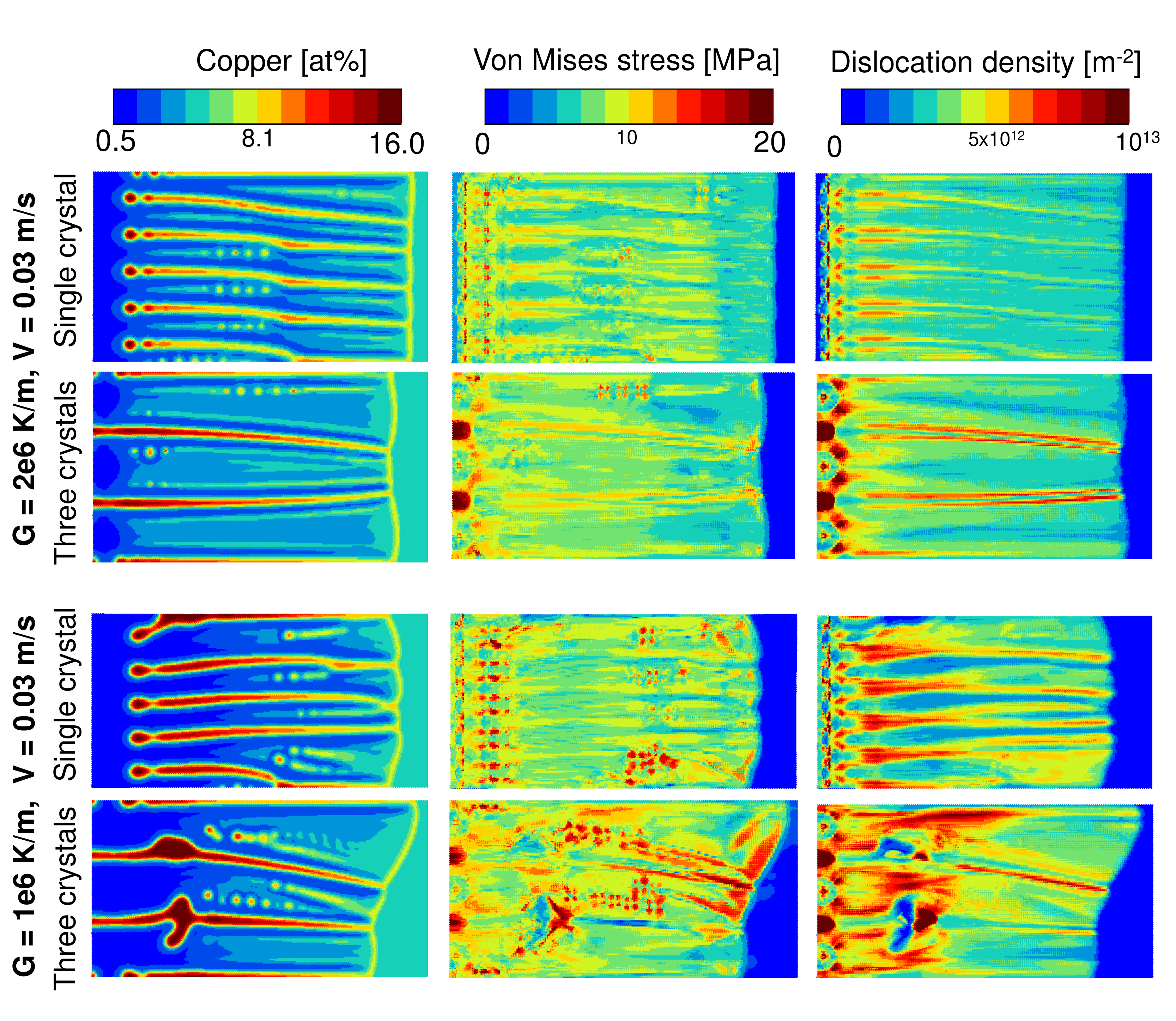}
\caption{Phase field -- crystal plasticity coupled simulation of Al--4.5at\%Cu rapid solidification for all four considered cases, showing the copper concentration, solidification shrinkage induced von Mises stresses, and the densities of statistically stored dislocations.}
\label{fig:PF_CP_Cs_mises_rhos_allCases}
\end{figure}

The concentration, von Mises stresses, and dislocation densities are shown for all four cases are shown in \autoref{fig:PF_CP_Cs_mises_rhos_allCases} before reaching the right system boundary. The smaller thermal gradient leads to larger heterogeneity in the solidification front, reflected as larger hot spots of von Mises stresses and dislocation densities. Moreover, the single crystal simulations lead to slightly lower dislocation densities than their multi-crystal counter-parts. This suggests that while the generation of large dislocation densities in rapid solidification is not inherently tied to competitive growth of neighboring crystals, the shrinkage stresses and the dislocation density distributions are magnified by grain-grain interactions. This stress magnification is consistent with experimental metal additive manufacturing (rapid solidification) of aluminium-copper alloys, where solidification cracking is observed more commonly along grain boundaries \cite{kotadia2021review}.

The average dislocation densities were calculated for the dislocation densities in \autoref{fig:PF_CP_Cs_mises_rhos_allCases}, where the last 130 nm from the right end of the system was excluded to avoid the liquid domain ahead of the solidification front. The average dislocation densities for each simulation case were as follows: 
4.1$\times$10$^{12}$ m$^{-2}$ for $G$ = 10$^6$ K/m single crystal,
4.9$\times$10$^{12}$ m$^{-2}$ for $G$ = 10$^6$ K/m three crystals,
3.6$\times$10$^{12}$ m$^{-2}$ for $G$ = 2$\times$10$^6$ K/m single crystal, and
4.1$\times$10$^{12}$ m$^{-2}$ for $G$ = 2$\times$10$^6$ K/m three crystals. These dislocation densities are summarized in Table \ref{table:dislocations} along with other simulation cases and experimentally obtained literature-given values, showing that the PF--CP simulations predict dislocation densities that are roughly 10-20 times smaller than those obtained experimentally.

There are several CP model parameters that are both challenging to determine accurately, and have a significant influence on the CP model quantitative predictions. 
In particular, dislocation climb can be come relevant during solidification, close to the alloy solidus. In this work, the effect of dislocation climb is phenomenologically included through the visco-plastic power-law-like strain rate rule (\autoref{eq:slip_rate}), parametrized in terms of the strain exponent $N$ and prefactor $K_\tau$. 
In our tests, decreasing the strain rate exponent $N$ in \autoref{eq:slip_rate} from 15 to 4 did not have a significant influence on the dislocation densities. Decreasing the dislocation hardening parameters $K_{forest}$ and $K_{coplan}$ in  \autoref{eq:dislocation_evolution} increased the average dislocation density levels to over 10$^14$ m$^{-2}$ which would better match the experimental results in \autoref{table:dislocations}; this parameter adjustment, however, resulted in an unrealistically large hardening behavior. In the future, crystal plasticity model parameters should be re-adjusted quantitatively to elevated temperature experiments or simulations with discrete dislocation dynamics (DDD) or MD.


\begin{table}
\caption{Dislocation densities in rapid solidification using various simulation methods, as well as measured in past experiments. The dislocation types are specified: geometrically necessary dislocations (GND), statistically stored dislocations (SSD), total dislocation content (Total). 
In MD simulations, the dislocation types were explicitly determined, where "Others" include stair-rod and Hirth type dislocations. In the past experiments listed, the measurement technique is indicated: elecron back-scatter diffraction (EBSD), X-ray diffraction (XRD), neutron diffraction (neutron diffr.), and transmission electron microscopy based line-intercept measurements (TEM line-inter.).}
\label{table:dislocations}
\begin{tabular}{cc l l l l l}
 \hline
 \multicolumn{6}{c}{Simulation results in this work} \\
&                        & Perfect  & Frank    & Shockley & Others & Unit \\ 
MD     & Pure Al         & 1.9      & 1.1      & 41       & 0.42   & 10$^{15}$ m$^{-2}$ \\
MD     & Al--1 at\%Cu    &  1.7     & 1.2      & 45       & 0.39   & 10$^{15}$ m$^{-2}$ \\ 
\hline 
PFC    & Pure Al, low V  & \multicolumn{4}{c}{0.1 (GND)}               & 10$^{15}$ m$^{-2}$ \\
PFC    & Pure Al, high V & \multicolumn{4}{c}{1.7 (GND)}               & 10$^{15}$ m$^{-2}$ \\ 
\hline
PF-CP  & Al--4.5at\%Cu   & \multicolumn{4}{c}{0.004-0.005 (SSD)}               & 10$^{15}$ m$^{-2}$ \\
\hline
\hline
 \multicolumn{6}{c}{Past experimental results} \\
\multicolumn{2}{c}{EBSD, AM Al-Cu \cite{hu2021}}  & \multicolumn{4}{c}{0.072 (GND)}               & 10$^{15}$ m$^{-2}$ \\
\multicolumn{2}{c}{XRD, Melt spun Al 7075 \cite{lin2015}}  & \multicolumn{4}{c}{0.057 (Total) }               & 10$^{15}$ m$^{-2}$ \\
\multicolumn{2}{c}{XRD, Melt spun Al 5083 \cite{lin2015_AlMg}}  & \multicolumn{4}{c}{0.14  (Total)}               & 10$^{15}$ m$^{-2}$ \\
\multicolumn{2}{c}{XRD, AM Al-Si-Mg \cite{zhang2021evolution}}  & \multicolumn{4}{c}{0.92 (Total)}               & 10$^{15}$ m$^{-2}$ \\
\multicolumn{2}{c}{Neutron diffr., AM Al-Si-Mg \cite{zhang2021multiscale}}  & \multicolumn{4}{c}{0.16-0.50 (Total)}               & 10$^{15}$ m$^{-2}$ \\
\multicolumn{2}{c}{TEM line-inter., AM Al-Si-Mg \cite{hadadzadeh2018columnar}}  & \multicolumn{4}{c}{0.11-0.31 (Total)}               & 10$^{15}$ m$^{-2}$ \\
\end{tabular}
\end{table}



\section{Discussion}
\label{section:discussion}
The MD simulations produced a relatively high density of Shockley partial dislocations forming around stacking faults that form around some of the grain boundaries, as shown in \autoref{fig:MD}. The density of intragrain dislocations (mostly perfect and Frank type) are remarkably close to the GND densities estimated in the rapid solidification PFC-AE simulations, as summarized in \autoref{table:dislocations}. These dislocation densities are 10-20 times larger than the listed experimentally measured dislocation densities \cite{lin2015,lin2015_AlMg,hu2021,hadadzadeh2018columnar,zhang2021evolution}. As discussed more thoroughly in the Introduction section, there are several sources of error in the experimental evaluation of dislocation densities that can, to some extent, explain the discrepancy between the MD and PFC--AE simulations and the experiments. 

The discrepancy can be related to the fact that some dislocations can annihilate during cool-down, and this cool-down is not simulated in this paper. 
Moreover, solidification velocities are somewhat higher in the MD and PFC--AE simulations than in typical AM or melt spun conditions, as indicated by the larger grain sizes in the experiments. Larger solidification velocities in the simulations are likely to pronounce the formation of dislocations. It should be noted that in this work, we did not perform a systematic investigation of solidification velocity versus dislocation density; we did, however, demonstrate a qualitative difference between "intermediate quench" and "large quench" regimes. The "intermediate quench" case was essentially free of intragrain GNDs with an overall lower average GND density, while the "large quench" case had significantly higher GND densities, visible intragrain lattice rotation, and trapped microvoid and smaller lattice defects.

Work is currently under way to match the PFC and PFC--AE models' kinetic time scales quantitatively to experimental conditions, which should help remedy this issue. Finally, the secondary hard phases in Al-Si-Mg alloy dislocation densities in Table \ref{table:dislocations}, from measurements in \cite{hadadzadeh2018columnar,zhang2021evolution}, are likely to have dislocations that are generated through thermal misfit between the FCC aluminium matrix and secondary hard phases that form during cool-down. These secondary phase effects are not accounted for in our simulations. 
In contrast to the MD and PFC--AE results, the PF--CP simulations lead to an order of magnitude smaller dislocation densities than what in the experiments in \autoref{table:dislocations}. This discrepancy can be related to 1) the uncertainty in the high temperature elastic and dislocation dynamics parameters used in the CP model; 2) lack of lattice curvature (orientation gradient) and GNDs in the CP model; 3) limited and indirect description of dislocation climb---a strongly thermally activated process---through the visoplastic flow rule parameters $K_\phi$ and $N$ in \autoref{eq:slip_rate}, which is known to be heavily thermally activated process.

While the MD was carried out in 3D, the PFC--AE and PF--CP simulations were in 2D. Performing the PF--CP simulations in 3D will lead to more complex grain morphologies and grain boundaries, which are likely to increase the heterogeneity due to solidification shrinkage and thereby the dislocation density levels. 
%
Currently, only pure aluminium was considered in the PFC simulations. Interaction between copper segregation and defect structures would be interesting to consider on diffusive time scales in PFC simulations, in order to better understand the emergence of pre-precipitates in rapid aluminium alloys. 
High quality mechanical properties for Al-Cu at temperatures close to the melting point are scattered and are a significant source of uncertainty for quantitative predictions in the PF--CP simulations. Some of these properties, especially elastic properties, could be evaluated with MD simulations.

\section{Conclusion}
In this paper, we presented a multiscale analysis of crystalline defect formation in pure aluminium and dilute Al-Cu alloys using molecular dynamics on smallest scales, phase field crystal based amplitude expansion (PFC--AE) simulations on intermediate scales, and a coupled phase field -- crystal plasticity (PF--CP) simulation scheme on largest scales. 

The mechanism of formation, merging, and healing of microvoids in rapid solidification of pure aluminium was demonstrated with PFC--AE simulations, being consistent with past experimental literature. 

The PFC-AE and MD simulations produced dislocation densities which were markedly close to each other, while being an order of magnitude larger than past experimentally measured dislocation densities. In contrast, the PF--CP framework lead to an order of magnitude smaller dislocation densities than in past experiments, where several potential sources of error were pointed out.

The presented multiscale framework can be used to investigate the formation of crystalline defects, as well as to perform calibration of uncertain parameters related to phenomenological dislocation evolution equations in the crystal plasticity models. As the submodels (MD, PFC--AE and PF--CP) underlying the platform presented continue to  evolve, it is expected that the quantitative prediction of dislocation densities with experiments will improve significantly.

\vskip6pt

\enlargethispage{20pt}





\section*{Funding}
TP, MH, and AL wish acknowledge the support of Academy of Finland through the HEADFORE project, Grant No. 333226, as well as CSC -- IT Center for Science, Finland, for computational resources. NP and PJ acknowledge the Natural Sciences and Engineering Research Council of Canada (NSERC) and the Canada Research Chairs (CRD) Program.



\bibliographystyle{unsrt}
\bibliography{bibliography}
\end{document}